\documentclass[aps,prl,showpacs,
  superscriptaddress,
  floatfix,
  twocolumn,
  amssymb,
  amsmath]{revtex4}

\pdfoutput=1

\usepackage{ifpdf}
\usepackage{graphicx}
\usepackage{dcolumn}
\usepackage{amsmath}
\usepackage{amsfonts}
\usepackage{amssymb}
\usepackage{bm}
\usepackage{color}
\usepackage{ulem}
\usepackage{longtable}

\begin{document}

\title{Epitaxial growth and magnetic properties of Sr$_{\mathbf{2}}$CrReO$_{\mathbf{6}}$ thin films}

\author{S.~Gepr\"{a}gs}
\email{Stephan.Gepraegs@wmi.badw.de}
 \affiliation{Walther-Mei{\ss}ner-Institut, Bayerische Akademie der Wissenschaften,
              D-85748 Garching, Germany}

\author{F.D.~Czeschka}
 \affiliation{Walther-Mei{\ss}ner-Institut, Bayerische Akademie der Wissenschaften,
              D-85748 Garching, Germany}

\author{M.~Opel}
 \affiliation{Walther-Mei{\ss}ner-Institut, Bayerische Akademie der Wissenschaften,
              D-85748 Garching, Germany}

\author{S.T.B.~Goennenwein}
 \affiliation{Walther-Mei{\ss}ner-Institut, Bayerische Akademie der Wissenschaften,
              D-85748 Garching, Germany}

\author{W.~Yu}
\affiliation{Institut f\"{u}r Anorganische Chemie, Universit\"{a}t Bonn, D-53117 Bonn,
Germany}

\author{W.~Mader}
\affiliation{Institut f\"{u}r Anorganische Chemie, Universit\"{a}t Bonn, D-53117 Bonn,
Germany}

\author{R.~Gross}
\email{Rudolf.Gross@wmi.badw.de}
 \affiliation{Walther-Mei{\ss}ner-Institut, Bayerische Akademie der Wissenschaften,
              D-85748 Garching, Germany}
 \affiliation{Physik-Department, Technische Universit\"{a}t M\"{u}nchen, D-85748
              Garching, Germany}

\date{\today}

\begin{abstract} % max. 100 words for APL
The double perovskite Sr$_2$CrReO$_6$ is an interesting material for
spintronics, showing ferrimagnetism up to 635\,K with a predicted high spin
polarization of $\approx86\%$. We fabricated Sr$_2$CrReO$_6$ epitaxial films by
pulsed laser deposition on (001)-oriented SrTiO$_3$ substrates. Phase-pure
films with optimum crystallographic and magnetic properties were obtained by
growing at a substrate temperature of $700^\circ$C in pure O$_2$ of $6.6\times
10^{-4}$\,mbar. The films are $c$-axis oriented, coherently strained, and show
less than 20\% anti-site defects. The magnetization curves reveal high
saturation magnetization of 0.8\,$\mu_{\rm B}$ per formula unit and high
coercivity of 1.1\,T, as well as a strong magnetic anisotropy.
\end{abstract}


\maketitle

\section{Introduction}\label{intro}

The ferrimagnetic double perovskites $A_2BB^\prime$O$_6$, with $B$ a magnetic
and $B^\prime$ a non-magnetic transition metal ion
\cite{Kobayashi1998,Serrate2007,Philipp2003b}, are attractive magnetic
materials. Of particular interest is Sr$_2$CrReO$_6$ (SCRO)
\cite{Kato2002,Asano2005}, as it exhibits a ferrimagnetic transition
temperature of $T_{\rm{C}}\approx635$\,K well above room temperature, in
combination with a quasi half-metallic band structure according to theoretical
calculations \cite{Vaitheeswaran2005}. This high $T_{\mathrm{C}}$ is strongly
correlated to a large induced magnetic spin moment at the non-magnetic Re site
\cite{Majewski2005a,Majewski2005b}. $T_{\mathrm{C}}$ can be further increased
by electron doping, as demonstrated e.g.~in Sr$_2$FeMoO$_6$, Sr$_2$CrWO$_6$, or
Sr$_2$CrOsO$_6$ by $5d$ band filling
\cite{Navarro2001,Geprags2006,Krockenberger2007}. The strong, long-range
magnetic interaction is attributed to a kinetic energy driven exchange process,
in which the hybridization between itinerant electrons on the $B^\prime$ sites
and the localized moments on the $B$ sites leads to an induced magnetic moment
of the non-magnetic $B^\prime$ ions \cite{Sarma2000,Fang2001}. Regarding device
applications, the strong influence of the growth conditions on the structural,
magnetic, and electronic properties of double perovskites is demanding.
In particular, it became evident that a key parameter to
understand the properties of double perovskites is the amount of anti-site
defects ($B$ ions on $B^{\prime}$ sites and vice versa). It was shown that the saturation magnetization $M_{\textrm{S}}$ 
decreases with increasing amount of anti-site defects
\cite{Balcells2001}. Therefore, a thorough understanding of the growth process
is a key prerequisite for obtaining thin films with optimum physical
properties. While the growth of Sr$_2$FeMoO$_6$ and Sr$_2$CrWO$_6$ thin films
has been investigated in
detail\cite{Westerburg2000,Philipp2001,Philipp2003a,Shinde2003}, little is
known about Sr$_2$CrReO$_6$ so far \cite{Asano2004,Asano2005,Asano2007}.

Here, we report on the epitaxial growth of $c$-axis oriented SCRO thin films on
(001) SrTiO$_3$ (STO) substrates, and discuss the effects of growth
temperature, growth atmosphere, and pressure on the film quality. We identify
optimal growth parameters leading to SCRO films with excellent crystalline
quality as well as magnetic properties comparable to bulk SCRO.

\section{Growth and crystallographic characterization}\label{fab}

We have used pulsed laser deposition (PLD) \cite{Klein1999} to grow the SCRO films
from a stoichiometric target, which was fabricated using an oxygen getter
control technique to ensure stoichiometry \cite{Yamamoto2000}. To identify the
optimal growth conditions, we have deposited films (i) in three different
atmospheres: O$_2$, Ar, and a mixture of Ar/O$_2$ (99/1 by volume), (ii) at
different pressures, and (iii) at different substrate temperatures
$450^\circ\mathrm{C}\le T_{\mathrm{G}}\le 900^\circ\mathrm{C}$. The structural
and  magnetic properties of the SCRO thin films were analyzed using in-situ
reflection high energy electron diffraction (RHEED), high resolution x-ray
diffraction (HRXRD), SQUID magnetometry, and high resolution transmission
electron microscopy (HRTEM). The film thickness was determined by x-ray
reflectometry.

Comparing samples grown in O$_2$, Ar, and Ar/O$_2$, we find that high-quality,
phase-pure SCRO films are obtained only in a growth atmosphere containing
oxygen. In particular, the presence of oxygen during the growth seems to be
essential to achieve Cr/Re sub-lattice ordering \cite{Venimadhav2004}. In the
following, we therefore focus on films grown in pure O$_2$ atmosphere of
$p_{\rm O_2}=6.6\times 10^{-4}$\,mbar. Higher pressure was found to lead to
parasitic phases, lower pressure to a reduced saturation magnetization. Another
important deposition parameter is the laser fluence $P_L$ on the target,
determining the energy of the ablated particles. We found no significant change
of the magnetic and crystalline quality upon varying $P_L$ from $1.0$ to
$2.5\,\textrm{J/cm}^{2}$. Moreover, the laser pulse repetition rate $f_L$ also
had no noticeable effect on the anti-site disorder in our SCRO films, in
contrast to what was found for Sr$_2$FeMoO$_6$ films \cite{Sanchez2004}. We
therefore used a $P_L = 2.0\,\textrm{J/cm}^{2}$ and $f_L = 2$\,Hz in all
subsequent growth runs.

%%%%%%%%%%%%%%%%% FIGURE 1: Growth %%%%%%%%%%%%%%%%%%%%%%%%%%%%%%%%%%%%%
\begin{figure}[tb]
   \centering
   \includegraphics [width=0.99\columnwidth,clip=]{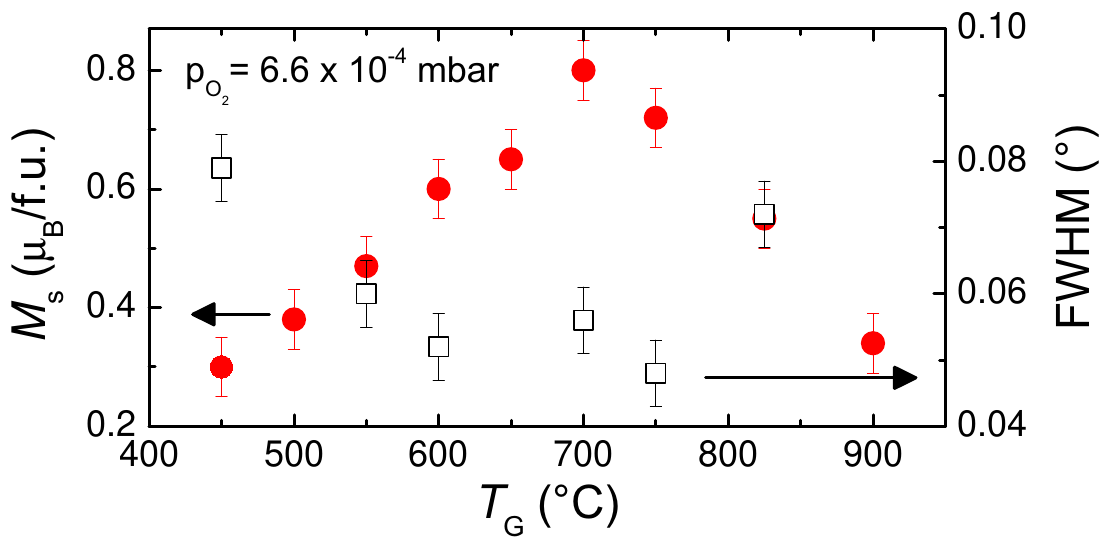}
   \caption{
Saturation magnetization $M_S$ measured at 25\,K and $\mu_{0}H=7~\mathrm{T}$ (full circles)
as well as FWHM of the rocking curve of the SCRO (004)
reflection (open squares) of single phase SCRO thin films grown
at $p_{\rm O_2}=6.6\times 10^{-4}$\,mbar plotted versus the substrate temperature
$T_{\mathrm{G}}$.
           }
 \label{Fig-FWHM}
\end{figure}
%%%%%%%%%%%%%%%%%%%%%%%%%%%%%%%%%%%%%%%%%%%%%%%%%%%%%%%%%%%%%%%%%%%%%%

The influence of the substrate temperature $T_{\rm G}$ on the structural and
magnetic properties of SCRO thin films is summarized in Fig.~\ref{Fig-FWHM}. It
is evident that the full width at half maximum (FWHM) of the rocking curve of
the SCRO (004) reflection stays almost constant on a low value of about
$0.05^\circ$ for $550^\circ\mathrm{C} \le T_{\mathrm{G}} \le
750^\circ\mathrm{C}$. Only at the lowest and highest $T_{\rm G}$, the FWHM
increases up to $0.08^\circ$. This indicates that SCRO thin films can be grown
with low mosaic spread in a large temperature window from $550^\circ$C to
$750^\circ$C. We note that the mosaic spread of the SCRO thin
films is close to that of the STO substrate (typically $0.01^\circ-0.02^\circ$). In contrast, the saturation magnetization $M_{\mathrm{S}}$ peaks around $T_{\rm
G}=700^\circ$C. As a strong correlation between $M_{\mathrm{S}}$ and the
sub-lattice order is often observed in ferrimagnetic double perovskites
\cite{Balcells2001}, Fig.~\ref{Fig-FWHM} suggests that SCRO films grown at
$T_{\rm G}\approx 700^\circ$C show a high degree of sub-lattice order and thus
a small fraction of anti-site defects. That is, SCRO thin films with optimal
structural and magnetic properties are obtained for $T_{\rm G}=700^\circ$C in a
pure O$_2$ atmosphere of $6.6 \times 10^{-4}$\,mbar.

%%%%%%%%%%%%%%%%% FIGURE 2: XRD %%%%%%%%%%%%%%%%%%%%%%%%%%%%%%%%%%%%%
\begin{figure}[tb]
    \centering
    \includegraphics [width=0.65\columnwidth,clip=]{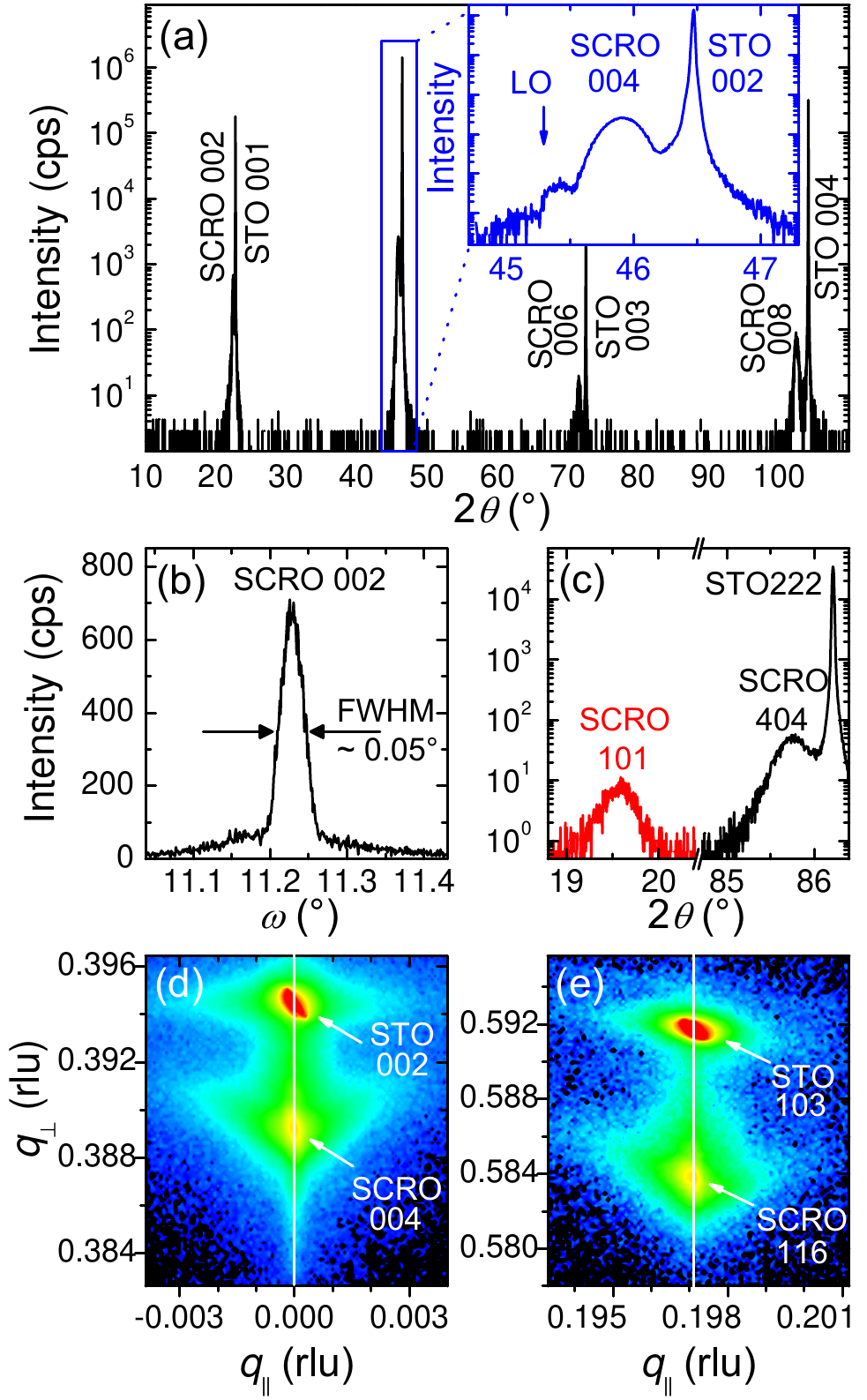}
    \caption{
        HRXRD of SCRO thin films grown at $700^\circ$C in $p_{\rm O_2}=6.6\times 10^{-4}$\,mbar.
        Panels (a)-(c) correspond to a SCRO film with
        a thickness of $d=31\,\mathrm{nm}$, panels (d) and (e) to a film with $d=65\,\mathrm{nm}$.
        (a) $\omega$-$2\theta$ scan showing Laue oscillations (LO),
        (b) rocking curve of the (002) reflection,
        (c) $\omega$-$2\theta$ scan of the superstructure (101) reflection,
        (d,e) reciprocal space maps around the (004) and (116) reflections.}
    \label{Fig-XRD}
\end{figure}
%%%%%%%%%%%%%%%%%%%%%%%%%%%%%%%%%%%%%%%%%%%%%%%%%%%%%%%%%%%%%%%%%%%%%%

Fig.~\ref{Fig-XRD} shows the structural properties achieved for epitaxial SCRO
films which were deposited at the optimum growth parameters described above.
The $\omega$-$2\theta$ scan reveals no crystalline parasitic phases. As shown
in the inset, the observation of Laue oscillations provides evidence for a
coherent film growth. Furthermore, the small FWHM of the rocking curve of the
SCRO (002) reflection of only $0.05^\circ$ demonstrates high crystalline
quality with low mosaic spread. The Cr/Re sub-lattice order of our SCRO films
can be inferred from the intensity of the superstructure (101) reflection shown
in Fig.~\ref{Fig-XRD}(c). This superstructure
peak has finite intensity only for a finite amount of ordering on the Cr/Re
sub-lattice. The amount of anti-site defects can be determined from the measured
intensity ratio of the (101) and (404) reflections. Comparing the measured
intensity to simulations, performed using the software
package TOPAS from Bruker company, the amount of anti-site defects is estimated to less
than 20\%. More detailed information about the
mosaicity and the strain state can be obtained from the reciprocal space maps
around the (004) and (116) reflections shown in Figs.~\ref{Fig-XRD}(d) and (e).
Clearly no $q_\|$-shift of the symmetric (004) and asymmetric (116) SCRO film
reflections with respect to the corresponding STO substrate reflections is
observed. This demonstrates pseudomorphic growth up to the maximum film
thickness of 65\,nm. The reciprocal space maps and the in-plane cube-on-cube
growth derived from an in-plane $\phi$-scan (not shown) reveal a tetragonal
symmetry of the SCRO thin films, with lattice parameters of
$a_{\textrm{SCRO}}=\sqrt{2} a_{\textrm{STO}}=5.523\textrm{\AA}$ and
$c_{\textrm{SCRO}}=7.90\textrm{\AA}$. Comparing the measured in-plane lattice
constant to the value $a_{\textrm{bulk}}=5.5206\textrm{\AA}$ of bulk SCRO
\cite{Serrate2007} shows that the tensile epitaxial coherency strain of the
SCRO films grown on STO substrates is negligibly small and the films can be
regarded as strain-free.

%%%%%%%%%%%%%%%%% FIGURE 3: TEM %%%%%%%%%%%%%%%%%%%%%%%%
\begin{figure}[tb]
   \centering
   \includegraphics [width=0.99\columnwidth,clip=]{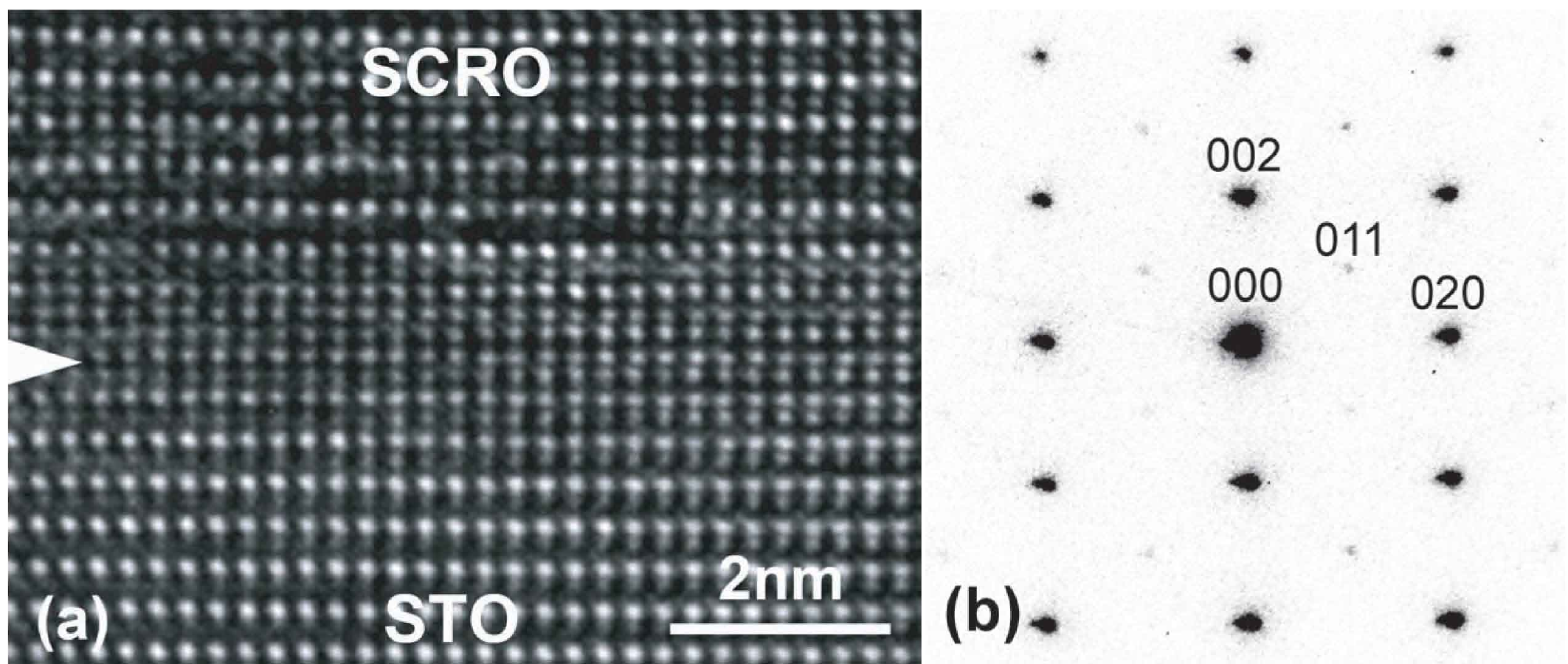}
    \caption{
(a) HRTEM micrograph of a 65\,nm thick, $c$-axis oriented SCRO film grown on (001) STO. The arrow marks the interface.
(b) Electron diffraction pattern of SCRO in $[100]$ orientation with odd-indexed super-lattice reflections.
           }
    \label{Fig-TEM}
\end{figure}

%%%%%%%%%%%%%%%%%%%%%%%%%%%%%%%%%%%%%%%%%%%%%%%%%%%%%%%%%%%%%%%%%%%%%%

The quality of the interface between the STO substrate and the SCRO film was
investigated by HRTEM. The bright-field HRTEM micrograph along the
$[110]$-direction of the STO substrate shown in Fig.~\ref{Fig-TEM}(a)
demonstrates the perfect epitaxial growth of SCRO on STO. Neither dislocations
nor structural defects could be detected. To obtain information on the Cr/Re
sub-lattice ordering, electron diffraction was performed, as shown in
Fig.~\ref{Fig-TEM}(b). After foil thickness determination from the intensity
ratio $I(020)/I(000)$, the anti-site defects of the Cr/Re sub-lattice can be
estimated to be less than 17\%, e.g. from the intensity ratio $I(011)/I(020)$.
Note that the (011) peak is a superstructure peak with vanishing
intensity for complete disorder on the Cr/Re sub-lattice. This
provides additional evidence for a highly ordered Cr/Re sub-lattice in our SCRO
films.

\section{Magnetic properties}\label{mag}

%%%%%%%%%%%%%%%%% FIGURE 4: Magnetic properties %%%%%%%%%%%%%%%%%%%%%%%%
\begin{figure}[b]
    \centering
    \includegraphics [width=0.99\columnwidth,clip=]{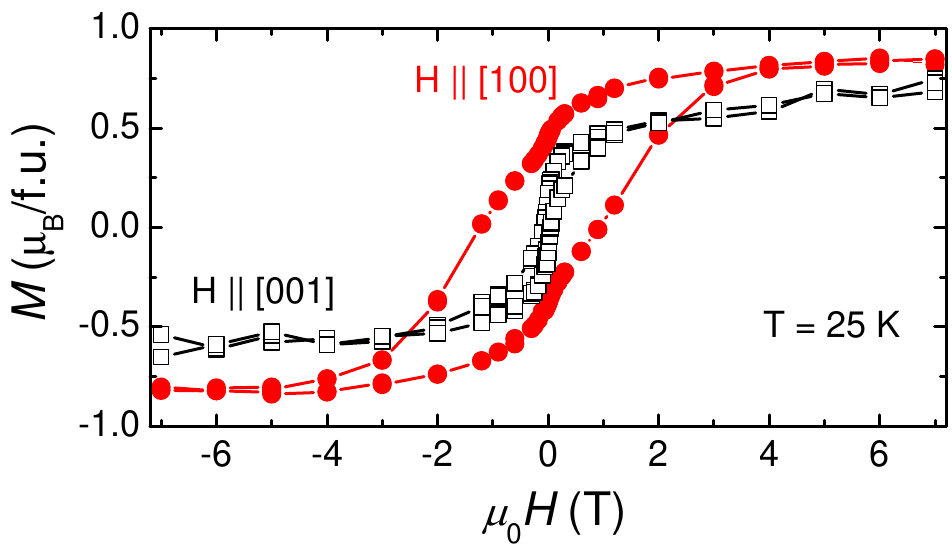}
    \caption{
Magnetization versus applied magnetic field curves of a 31\,nm thick SCRO film
measured by SQUID magnetometry at $T=25$\,K for $H$ parallel (full circles) and
perpendicular (open squares) to the film plane.
            }
    \label{Fig-SQUID}
\end{figure}
%%%%%%%%%%%%%%%%%%%%%%%%%%%%%%%%%%%%%%%%%%%%%%%%%%%%%%%%%%%%%%%%%%%%%%

We finally address the magnetic properties of the SCRO films.
Fig.~\ref{Fig-SQUID} exemplarily shows the $M(H)$ curves of a 31\,nm thick SCRO
film recorded at 25\,K. For the external magnetic field applied in-plane along
$\left[100\right]$, the measured saturation magnetization
$M_{\mathrm{S}}=0.8\,\mu_{\rm B}/\textrm{formula unit (f.u.)}$ is in good
agreement with $M_{\mathrm{S}}=0.89\,\mu_{\rm B}/\mathrm{f.u.}$ of our
polycrystalline target material and the literature values ranging between 0.8
and $0.9\,\mu_{\rm B}/\mathrm{f.u.}$ at low temperature
\cite{Kato2002,Asano2004}. Recent model calculations including spin-orbit
coupling \cite{Vaitheeswaran2005} predicted $M_{\mathrm{S}}=1.28\,\mu_{\rm
B}$/f.u. for perfect sub-lattice order. Assuming that $M_{\mathrm{S}}$
decreases about linearly with increasing sub-lattice disorder, the measured
amount of anti-site defects of less than 20\% translates into an expected
saturation magnetization larger than $0.77\,\mu_{\rm B}$/f.u., again in good
agreement with the measured value. The measured $M(H)$ curves also reveal a
very large coercive field $\mu_{0}H_{\mathrm{c}} = 1.1\,\mathrm{T}$. This value
again agrees well with literature data ranging from 1 to 1.5\,T at 4.2\,K
\cite{Kato2002,Asano2004}, demonstrating that the SCRO compound is a hard
ferromagnet. However, in contrast to a previous study \cite{Asano2004} we
observe a strong magnetic anisotropy in our SCRO films. This is evident from
the $M(H)$ hysteresis loop with $H$ oriented perpendicular to the film plane.
For this field orientation, the magnetization loop is rectangular-shaped, with
a much lower coercive field $\mu_{0}H_{\mathrm{c}} = 62\,\mathrm{mT}$.
Furthermore, the field at which the hysteresis loop closes is also reduced to
about 0.7\,T, as compared to more than 3\,T for the magnetic field applied in
the film plane. We note that the apparent dependence of $M_{\mathrm{S}}$ on the
field orientation can be attributed to the large saturation field of SCRO,
making the unambiguous determination of the diamagnetic contribution of the
substrate difficult. Additionally, the large spin-orbit coupling in SCRO is
expected to result in strong magnetic anisotropy.

\section{Conclusions}

In conclusion, we have identified the optimal parameters for the epitaxial
growth of $c$-axis oriented SCRO films on (001) STO by pulsed laser deposition.
The films have excellent crystalline quality with a high degree of Cr/Re
sub-lattice order. The FWHM of the rocking curves is as small as $0.05^\circ$,
$M_{\mathrm{S}}$ reaches $0.8\,\mu_{\rm B}$/f.u., and the coercivity is 1.1\,T.
These parameters make SCRO thin films an outstanding ferrimagnetic double
perovskite for spintronic applications.

Financial support by the DFG via the priority programs 1157 and 1285 (project
nos. GR 1132/13 and 1132/14), GO 944/3-1, and the Excellence Cluster "Nanosystems Initiative
Munich (NIM)" are gratefully acknowledged.

\clearpage


\small

\begin{thebibliography}{24}
\expandafter\ifx\csname natexlab\endcsname\relax\def\natexlab#1{#1}\fi
\expandafter\ifx\csname bibnamefont\endcsname\relax
  \def\bibnamefont#1{#1}\fi
\expandafter\ifx\csname bibfnamefont\endcsname\relax
  \def\bibfnamefont#1{#1}\fi
\expandafter\ifx\csname citenamefont\endcsname\relax
  \def\citenamefont#1{#1}\fi
\expandafter\ifx\csname url\endcsname\relax
  \def\url#1{\texttt{#1}}\fi
\expandafter\ifx\csname urlprefix\endcsname\relax\def\urlprefix{URL }\fi
\providecommand{\bibinfo}[2]{#2}
\providecommand{\eprint}[2][]{\url{#2}}

\bibitem[{\citenamefont{Kobayashi et~al.}(1998)\citenamefont{Kobayashi, Kimura,
  Sawada, Terakura, and Tokura}}]{Kobayashi1998}
\bibinfo{author}{\bibfnamefont{K.-I.} \bibnamefont{Kobayashi}},
  \bibinfo{author}{\bibfnamefont{T.}~\bibnamefont{Kimura}},
  \bibinfo{author}{\bibfnamefont{H.}~\bibnamefont{Sawada}},
  \bibinfo{author}{\bibfnamefont{K.}~\bibnamefont{Terakura}}, \bibnamefont{and}
  \bibinfo{author}{\bibfnamefont{Y.}~\bibnamefont{Tokura}},
  \bibinfo{journal}{Nature} \textbf{\bibinfo{volume}{395}},
  \bibinfo{pages}{677} (\bibinfo{year}{1998}).

\bibitem[{\citenamefont{Serrate et~al.}(2007)\citenamefont{Serrate, Teresa, and
  Ibarra}}]{Serrate2007}
\bibinfo{author}{\bibfnamefont{D.}~\bibnamefont{Serrate}},
  \bibinfo{author}{\bibfnamefont{J.~D.} \bibnamefont{Teresa}}, \bibnamefont{and}
  \bibinfo{author}{\bibfnamefont{M.}~\bibnamefont{Ibarra}},
  \bibinfo{journal}{J. Phys.: Cond. Matt.} \textbf{\bibinfo{volume}{19}},
  \bibinfo{pages}{023201} (\bibinfo{year}{2007}).

\bibitem[{\citenamefont{Philipp
  et~al.}(2003{\natexlab{a}})\citenamefont{Philipp, Majewski, Alff, Erb, Gross,
  Graf, Brandt, Simon, Walther, Mader et~al.}}]{Philipp2003b}
\bibinfo{author}{\bibfnamefont{J.~B.} \bibnamefont{Philipp}},
  \bibinfo{author}{\bibfnamefont{P.}~\bibnamefont{Majewski}},
  \bibinfo{author}{\bibfnamefont{L.}~\bibnamefont{Alff}},
  \bibinfo{author}{\bibfnamefont{A.}~\bibnamefont{Erb}},
  \bibinfo{author}{\bibfnamefont{R.}~\bibnamefont{Gross}},
  \bibinfo{author}{\bibfnamefont{T.}~\bibnamefont{Graf}},
  \bibinfo{author}{\bibfnamefont{M.~S.} \bibnamefont{Brandt}},
  \bibinfo{author}{\bibfnamefont{J.}~\bibnamefont{Simon}},
  \bibinfo{author}{\bibfnamefont{T.}~\bibnamefont{Walther}},
  \bibinfo{author}{\bibfnamefont{W.}~\bibnamefont{Mader}},
  \bibinfo{author}{\bibfnamefont{D.}~\bibnamefont{Topwal}}, \bibnamefont{and}
  \bibinfo{author}{\bibfnamefont{D.~D.}~\bibnamefont{Sarma}},
  \bibinfo{journal}{Phys. Rev. B}
  \textbf{\bibinfo{volume}{68}}, \bibinfo{pages}{144431}
  (\bibinfo{year}{2003}{\natexlab{a}}).

\bibitem[{\citenamefont{Kato et~al.}(2002)\citenamefont{Kato, Okuda, Okimoto,
  Tomioka, Takenoya, Ohkubo, Kawasaki, and Tokura}}]{Kato2002}
\bibinfo{author}{\bibfnamefont{H.}~\bibnamefont{Kato}},
  \bibinfo{author}{\bibfnamefont{T.}~\bibnamefont{Okuda}},
  \bibinfo{author}{\bibfnamefont{Y.}~\bibnamefont{Okimoto}},
  \bibinfo{author}{\bibfnamefont{Y.}~\bibnamefont{Tomioka}},
  \bibinfo{author}{\bibfnamefont{Y.}~\bibnamefont{Takenoya}},
  \bibinfo{author}{\bibfnamefont{A.}~\bibnamefont{Ohkubo}},
  \bibinfo{author}{\bibfnamefont{M.}~\bibnamefont{Kawasaki}}, \bibnamefont{and}
  \bibinfo{author}{\bibfnamefont{Y.}~\bibnamefont{Tokura}},
  \bibinfo{journal}{Appl. Phys. Lett.} \textbf{\bibinfo{volume}{81}},
  \bibinfo{pages}{328} (\bibinfo{year}{2002}).

\bibitem[{\citenamefont{Asano et~al.}(2005)\citenamefont{Asano, Koduka, Imaeda,
  Sugiyama, and Matsui}}]{Asano2005}
\bibinfo{author}{\bibfnamefont{H.}~\bibnamefont{Asano}},
  \bibinfo{author}{\bibfnamefont{N.}~\bibnamefont{Koduka}},
  \bibinfo{author}{\bibfnamefont{K.}~\bibnamefont{Imaeda}},
  \bibinfo{author}{\bibfnamefont{M.}~\bibnamefont{Sugiyama}}, \bibnamefont{and}
  \bibinfo{author}{\bibfnamefont{M.}~\bibnamefont{Matsui}},
  \bibinfo{journal}{IEEE Trans. Magn.} \textbf{\bibinfo{volume}{41}},
  \bibinfo{pages}{2811} (\bibinfo{year}{2005}).

\bibitem[{\citenamefont{Vaitheeswaran et~al.}(2005)\citenamefont{Vaitheeswaran,
  Kanchana, and Delin}}]{Vaitheeswaran2005}
\bibinfo{author}{\bibfnamefont{G.}~\bibnamefont{Vaitheeswaran}},
  \bibinfo{author}{\bibfnamefont{V.}~\bibnamefont{Kanchana}}, \bibnamefont{and}
  \bibinfo{author}{\bibfnamefont{A.}~\bibnamefont{Delin}},
  \bibinfo{journal}{Appl. Phys. Lett.} \textbf{\bibinfo{volume}{86}},
  \bibinfo{eid}{032513} (\bibinfo{year}{2005}).

\bibitem[{\citenamefont{Majewski
  et~al.}(2005{\natexlab{a}})\citenamefont{Majewski, Gepr\"{a}gs, Boger, Opel,
  Erb, Gross, Vaitheeswaran, Kanchana, Delin, Wilhelm et~al.}}]{Majewski2005a}
\bibinfo{author}{\bibfnamefont{P.}~\bibnamefont{Majewski}},
  \bibinfo{author}{\bibfnamefont{S.}~\bibnamefont{Gepr\"{a}gs}},
  \bibinfo{author}{\bibfnamefont{A.}~\bibnamefont{Boger}},
  \bibinfo{author}{\bibfnamefont{M.}~\bibnamefont{Opel}},
  \bibinfo{author}{\bibfnamefont{A.}~\bibnamefont{Erb}},
  \bibinfo{author}{\bibfnamefont{R.}~\bibnamefont{Gross}},
  \bibinfo{author}{\bibfnamefont{G.}~\bibnamefont{Vaitheeswaran}},
  \bibinfo{author}{\bibfnamefont{V.}~\bibnamefont{Kanchana}},
  \bibinfo{author}{\bibfnamefont{A.}~\bibnamefont{Delin}},
  \bibinfo{author}{\bibfnamefont{F.}~\bibnamefont{Wilhelm}},
  \bibinfo{author}{\bibfnamefont{A.}~\bibnamefont{Rogalev}}, \bibnamefont{and}
  \bibinfo{author}{\bibfnamefont{L.}~\bibnamefont{Alff}},
  \bibinfo{journal}{Phys. Rev. B}
  \textbf{\bibinfo{volume}{72}}, \bibinfo{eid}{132402}
  (\bibinfo{year}{2005}{\natexlab{a}}).

\bibitem[{\citenamefont{Majewski
  et~al.}(2005{\natexlab{b}})\citenamefont{Majewski, Gepr\"{a}gs, Sanganas,
  Opel, Gross, Wilhelm, Rogalev, and Alff}}]{Majewski2005b}
\bibinfo{author}{\bibfnamefont{P.}~\bibnamefont{Majewski}},
  \bibinfo{author}{\bibfnamefont{S.}~\bibnamefont{Gepr\"{a}gs}},
  \bibinfo{author}{\bibfnamefont{O.}~\bibnamefont{Sanganas}},
  \bibinfo{author}{\bibfnamefont{M.}~\bibnamefont{Opel}},
  \bibinfo{author}{\bibfnamefont{R.}~\bibnamefont{Gross}},
  \bibinfo{author}{\bibfnamefont{F.}~\bibnamefont{Wilhelm}},
  \bibinfo{author}{\bibfnamefont{A.}~\bibnamefont{Rogalev}}, \bibnamefont{and}
  \bibinfo{author}{\bibfnamefont{L.}~\bibnamefont{Alff}},
  \bibinfo{journal}{Appl. Phys. Lett.} \textbf{\bibinfo{volume}{87}},
  \bibinfo{eid}{202503} (\bibinfo{year}{2005}{\natexlab{b}}).

\bibitem[{\citenamefont{Navarro et~al.}(2001)\citenamefont{Navarro, Frontera,
  Balcells, Martinez, and Fontcuberta}}]{Navarro2001}
\bibinfo{author}{\bibfnamefont{J.}~\bibnamefont{Navarro}},
  \bibinfo{author}{\bibfnamefont{C.}~\bibnamefont{Frontera}},
  \bibinfo{author}{\bibfnamefont{L.}~\bibnamefont{Balcells}},
  \bibinfo{author}{\bibfnamefont{B.}~\bibnamefont{Martinez}}, \bibnamefont{and}
  \bibinfo{author}{\bibfnamefont{J.}~\bibnamefont{Fontcuberta}},
  \bibinfo{journal}{Phys. Rev. B} \textbf{\bibinfo{volume}{64}},
  \bibinfo{eid}{092411} (\bibinfo{year}{2001}).

\bibitem[{\citenamefont{Gepr\"{a}gs et~al.}(2006)\citenamefont{Gepr\"{a}gs,
  Majewski, Gross, Ritter, and Alff}}]{Geprags2006}
\bibinfo{author}{\bibfnamefont{S.}~\bibnamefont{Gepr\"{a}gs}},
  \bibinfo{author}{\bibfnamefont{P.}~\bibnamefont{Majewski}},
  \bibinfo{author}{\bibfnamefont{R.}~\bibnamefont{Gross}},
  \bibinfo{author}{\bibfnamefont{C.}~\bibnamefont{Ritter}}, \bibnamefont{and}
  \bibinfo{author}{\bibfnamefont{L.}~\bibnamefont{Alff}}
  \bibinfo{journal}{J. Appl. Phys.} \textbf{\bibinfo{volume}{99}}, \bibinfo{pages}{08J102}
  (\bibinfo{year}{2006}{\natexlab{b}}).

\bibitem[{\citenamefont{Krockenberger et~al.}(2007)\citenamefont{Krockenberger,
  Mogare, Reehuis, Tovar, Jansen, Vaitheeswaran, Kanchana, Bultmark, Delin,
  Wilhelm et~al.}}]{Krockenberger2007}
\bibinfo{author}{\bibfnamefont{Y.}~\bibnamefont{Krockenberger}},
  \bibinfo{author}{\bibfnamefont{K.}~\bibnamefont{Mogare}},
  \bibinfo{author}{\bibfnamefont{M.}~\bibnamefont{Reehuis}},
  \bibinfo{author}{\bibfnamefont{M.}~\bibnamefont{Tovar}},
  \bibinfo{author}{\bibfnamefont{M.}~\bibnamefont{Jansen}},
  \bibinfo{author}{\bibfnamefont{G.}~\bibnamefont{Vaitheeswaran}},
  \bibinfo{author}{\bibfnamefont{V.}~\bibnamefont{Kanchana}},
  \bibinfo{author}{\bibfnamefont{F.}~\bibnamefont{Bultmark}},
  \bibinfo{author}{\bibfnamefont{A.}~\bibnamefont{Delin}},
  \bibinfo{author}{\bibfnamefont{F.}~\bibnamefont{Wilhelm}},
  \bibinfo{author}{\bibfnamefont{A.}~\bibnamefont{Rogalev}},
  \bibinfo{author}{\bibfnamefont{A.}~\bibnamefont{Winkler}}, \bibnamefont{and}
  \bibinfo{author}{\bibfnamefont{L.}~\bibnamefont{Alff}},
  \bibinfo{journal}{Phys. Rev. B}
  \textbf{\bibinfo{volume}{75}}, \bibinfo{eid}{020404} (\bibinfo{year}{2007}).

\bibitem[{\citenamefont{Sarma et~al.}(2000)\citenamefont{Sarma, Mahadevan,
  Saha-Dasgupta, Ray, and Kumar}}]{Sarma2000}
\bibinfo{author}{\bibfnamefont{D.~D.} \bibnamefont{Sarma}},
  \bibinfo{author}{\bibfnamefont{P.}~\bibnamefont{Mahadevan}},
  \bibinfo{author}{\bibfnamefont{T.}~\bibnamefont{Saha-Dasgupta}},
  \bibinfo{author}{\bibfnamefont{S.}~\bibnamefont{Ray}}, \bibnamefont{and}
  \bibinfo{author}{\bibfnamefont{A.}~\bibnamefont{Kumar}},
  \bibinfo{journal}{Phys. Rev. Lett.} \textbf{\bibinfo{volume}{85}},
  \bibinfo{pages}{2549} (\bibinfo{year}{2000}).

\bibitem[{\citenamefont{Fang et~al.}(2001)\citenamefont{Fang, Terakura, and
  Kanamori}}]{Fang2001}
\bibinfo{author}{\bibfnamefont{Z.}~\bibnamefont{Fang}},
  \bibinfo{author}{\bibfnamefont{K.}~\bibnamefont{Terakura}}, \bibnamefont{and}
  \bibinfo{author}{\bibfnamefont{J.}~\bibnamefont{Kanamori}},
  \bibinfo{journal}{Phys. Rev. B} \textbf{\bibinfo{volume}{63}},
  \bibinfo{eid}{180407} (\bibinfo{year}{2001}).

\bibitem[{\citenamefont{Balcells et~al.}(2001)\citenamefont{Balcells, Navarro,
  Bibes, Roig, Martinez, and Fontcuberta}}]{Balcells2001}
\bibinfo{author}{\bibfnamefont{L.}~\bibnamefont{Balcells}},
  \bibinfo{author}{\bibfnamefont{J.}~\bibnamefont{Navarro}},
  \bibinfo{author}{\bibfnamefont{M.}~\bibnamefont{Bibes}},
  \bibinfo{author}{\bibfnamefont{A.}~\bibnamefont{Roig}},
  \bibinfo{author}{\bibfnamefont{B.}~\bibnamefont{Martinez}}, \bibnamefont{and}
  \bibinfo{author}{\bibfnamefont{J.}~\bibnamefont{Fontcuberta}},
  \bibinfo{journal}{Appl. Phys. Lett.} \textbf{\bibinfo{volume}{78}},
  \bibinfo{pages}{781} (\bibinfo{year}{2001}).

\bibitem[{\citenamefont{Westerburg et~al.}(2000)\citenamefont{Westerburg,
  Reisinger, and Jakob}}]{Westerburg2000}
\bibinfo{author}{\bibfnamefont{W.}~\bibnamefont{Westerburg}},
  \bibinfo{author}{\bibfnamefont{D.}~\bibnamefont{Reisinger}}, \bibnamefont{and}
  \bibinfo{author}{\bibfnamefont{G.}~\bibnamefont{Jakob}},
  \bibinfo{journal}{Phys. Rev. B} \textbf{\bibinfo{volume}{62}},
  \bibinfo{pages}{R767} (\bibinfo{year}{2000}).

\bibitem[{\citenamefont{Philipp et~al.}(2001)\citenamefont{Philipp, Reisinger,
  Schonecke, Marx, Erb, Alff, Gross, and Klein}}]{Philipp2001}
\bibinfo{author}{\bibfnamefont{J.~B.} \bibnamefont{Philipp}},
  \bibinfo{author}{\bibfnamefont{D.}~\bibnamefont{Reisinger}},
  \bibinfo{author}{\bibfnamefont{M.}~\bibnamefont{Schonecke}},
  \bibinfo{author}{\bibfnamefont{A.}~\bibnamefont{Marx}},
  \bibinfo{author}{\bibfnamefont{A.}~\bibnamefont{Erb}},
  \bibinfo{author}{\bibfnamefont{L.}~\bibnamefont{Alff}},
  \bibinfo{author}{\bibfnamefont{R.}~\bibnamefont{Gross}}, \bibnamefont{and}
  \bibinfo{author}{\bibfnamefont{J.}~\bibnamefont{Klein}},
  \bibinfo{journal}{Appl. Phys. Lett.} \textbf{\bibinfo{volume}{79}},
  \bibinfo{pages}{3654} (\bibinfo{year}{2001}).

\bibitem[{\citenamefont{Philipp
  et~al.}(2003{\natexlab{b}})\citenamefont{Philipp, Reisinger, Schonecke, Opel,
  Marx, Erb, Alff, and Gross}}]{Philipp2003a}
\bibinfo{author}{\bibfnamefont{J.~B.} \bibnamefont{Philipp}},
  \bibinfo{author}{\bibfnamefont{D.}~\bibnamefont{Reisinger}},
  \bibinfo{author}{\bibfnamefont{M.}~\bibnamefont{Schonecke}},
  \bibinfo{author}{\bibfnamefont{M.}~\bibnamefont{Opel}},
  \bibinfo{author}{\bibfnamefont{A.}~\bibnamefont{Marx}},
  \bibinfo{author}{\bibfnamefont{A.}~\bibnamefont{Erb}},
  \bibinfo{author}{\bibfnamefont{L.}~\bibnamefont{Alff}}, \bibnamefont{and}
  \bibinfo{author}{\bibfnamefont{R.}~\bibnamefont{Gross}},
  \bibinfo{journal}{J. Appl. Phys.} \textbf{\bibinfo{volume}{93}}, \bibinfo{pages}{6853}
  (\bibinfo{year}{2003}{\natexlab{b}}).

\bibitem[{\citenamefont{Shinde et~al.}(2003)\citenamefont{Shinde, Ogale,
  Greene, Venkatesan, Tsoi, Cheong, and Millis}}]{Shinde2003}
\bibinfo{author}{\bibfnamefont{S.~R.} \bibnamefont{Shinde}},
  \bibinfo{author}{\bibfnamefont{S.~B.} \bibnamefont{Ogale}},
  \bibinfo{author}{\bibfnamefont{R.~L.} \bibnamefont{Greene}},
  \bibinfo{author}{\bibfnamefont{T.}~\bibnamefont{Venkatesan}},
  \bibinfo{author}{\bibfnamefont{K.}~\bibnamefont{Tsoi}},
  \bibinfo{author}{\bibfnamefont{S.-W.} \bibnamefont{Cheong}}, \bibnamefont{and}
  \bibinfo{author}{\bibfnamefont{A.~J.} \bibnamefont{Millis}},
  \bibinfo{journal}{J. Appl. Phys.}
  \textbf{\bibinfo{volume}{93}}, \bibinfo{pages}{1605} (\bibinfo{year}{2003}).

\bibitem[{\citenamefont{Asano et~al.}(2004)\citenamefont{Asano, Kozuka,
  Tsuzuki, and Matsui}}]{Asano2004}
\bibinfo{author}{\bibfnamefont{H.}~\bibnamefont{Asano}},
  \bibinfo{author}{\bibfnamefont{N.}~\bibnamefont{Kozuka}},
  \bibinfo{author}{\bibfnamefont{A.}~\bibnamefont{Tsuzuki}}, \bibnamefont{and}
  \bibinfo{author}{\bibfnamefont{M.}~\bibnamefont{Matsui}},
  \bibinfo{journal}{Appl. Phys. Lett.} \textbf{\bibinfo{volume}{85}},
  \bibinfo{pages}{263} (\bibinfo{year}{2004}).

\bibitem[{\citenamefont{Asano et~al.}(2007)\citenamefont{Asano, Koduka,
  Takahashi, and Matsui}}]{Asano2007}
\bibinfo{author}{\bibfnamefont{H.}~\bibnamefont{Asano}},
  \bibinfo{author}{\bibfnamefont{N.}~\bibnamefont{Koduka}},
  \bibinfo{author}{\bibfnamefont{Y.}~\bibnamefont{Takahashi}}, \bibnamefont{and}
  \bibinfo{author}{\bibfnamefont{M.}~\bibnamefont{Matsui}},
  \bibinfo{journal}{J. Magn. Magn. Mat.} \textbf{\bibinfo{volume}{310}},
  \bibinfo{pages}{2174} (\bibinfo{year}{2007}).

\bibitem[{\citenamefont{Klein et~al.}(1999)\citenamefont{Klein, H\"{o}fener, Alff,
  and Gross}}]{Klein1999}
\bibinfo{author}{\bibfnamefont{J.}~\bibnamefont{Klein}},
  \bibinfo{author}{\bibfnamefont{C.}~\bibnamefont{H\"{o}fener}},
  \bibinfo{author}{\bibfnamefont{L.}~\bibnamefont{Alff}}, \bibnamefont{and}
  \bibinfo{author}{\bibfnamefont{R.}~\bibnamefont{Gross}},
  \bibinfo{journal}{Superc. Sci. Techn.} \textbf{\bibinfo{volume}{12}},
  \bibinfo{pages}{1023} (\bibinfo{year}{1999}).

\bibitem[{\citenamefont{Yamamoto et~al.}(2000)\citenamefont{Yamamoto,
  Liimatainen, Lind\'{e}n, Karppinen, and Yamauchi}}]{Yamamoto2000}
\bibinfo{author}{\bibfnamefont{T.}~\bibnamefont{Yamamoto}},
  \bibinfo{author}{\bibfnamefont{J.}~\bibnamefont{Liimatainen}},
  \bibinfo{author}{\bibfnamefont{J.}~\bibnamefont{Lind\'{e}n}},
  \bibinfo{author}{\bibfnamefont{M.}~\bibnamefont{Karppinen}},
  \bibnamefont{and} \bibinfo{author}{\bibfnamefont{H.}~\bibnamefont{Yamauchi}},
  \bibinfo{journal}{J. Mat. Chem.} \textbf{\bibinfo{volume}{10}},
  \bibinfo{pages}{2342} (\bibinfo{year}{2000}).

\bibitem[{\citenamefont{Venimadhav et~al.}(2006)\citenamefont{Venimadhav, Sher,
  Attfield, and Blamire}}]{Venimadhav2004}
\bibinfo{author}{\bibfnamefont{A.}~\bibnamefont{Venimadhav}},
  \bibinfo{author}{\bibfnamefont{F.}~\bibnamefont{Sher}},
  \bibinfo{author}{\bibfnamefont{J.~P.}~\bibnamefont{Attfield}}, \bibnamefont{and}
  \bibinfo{author}{\bibfnamefont{M.~G.}~\bibnamefont{Blamire}},
  \bibinfo{journal}{Sol. State Comm.} \textbf{\bibinfo{volume}{138}},
  \bibinfo{pages}{314} (\bibinfo{year}{2006}).

\bibitem[{\citenamefont{Sanchez et~al.}(2004)\citenamefont{Sanchez,
  Garcia-Hernandez, Auth, and Jakob}}]{Sanchez2004}
\bibinfo{author}{\bibfnamefont{D.}~\bibnamefont{Sanchez}},
  \bibinfo{author}{\bibfnamefont{M.}~\bibnamefont{Garcia-Hernandez}},
  \bibinfo{author}{\bibfnamefont{N.}~\bibnamefont{Auth}}, \bibnamefont{and}
  \bibinfo{author}{\bibfnamefont{G.}~\bibnamefont{Jakob}},
  \bibinfo{journal}{J. Appl. Phys.} \textbf{\bibinfo{volume}{96}}, \bibinfo{pages}{2736}
  (\bibinfo{year}{2004}).

\end{thebibliography}
\end{document}